\documentclass[twocolumn,showpacs,preprintnumbers,amsmath,amssymb]{revtex4}

\usepackage{graphicx}
\usepackage{amsmath,amsfonts}
\begin{document}

\author{David Kleinhans}
\author{Rudolf Friedrich}
\affiliation{Westf{\"a}lische Wilhelms-Universit{\"a}t M{\"u}nster, D-48149 M{\"u}nster, Germany}

\title{Maximum Likelihood Estimation of Drift and Diffusion Functions}
\date{\today}


\begin{abstract}
The maximum likelihood approach is adapted to the problem
of  estimation
of drift and diffusion functions of stochastic processes 
from measured time series. 
We reconcile a previously devised iterative procedure
\cite{Kleinhans05} and put the application of the method
on a firm theoretical basis. 
\end{abstract}

\maketitle

\section{Introduction}
Complex systems of physics, chemistry, and biology 
are composed of a huge number
of microscopic
subsystems interacting on a fast time scale. Self organized behaviour
may arise on a macroscopic length and time scale which can be described by
suitably defined order parameters. The microscopic degree's of freedom, 
however, show up in terms of fast temporal variations which effectively can be
treated as random fluctuations \cite{Haken:Synergetics}.
The adequate description of such systems viewed from a macroscopic 
perspective  
are Langevin equations, which contain a deterministic part described by 
the drift vector and a fluctuating part whose impact on the dynamics is
quantified by a diffusion matrix \cite{Risken,Gardiner}.

Recently, a procedure has been proposed that allows for a direct estimation
of these quantities and, hence, of the stochastic 
dynamics from measured data
\cite{Siegert98}. This procedure has provided a deeper insight to a
broad class of systems, especially in the field of \emph{life
sciences} \cite{Kriso02,Siegert98,Kuusela04}. Moreover, also
turbulence research has greatly 
benefited from this procedure \cite{Friedrich97}.

However, the procedure is based on the estimation of conditional
moments in the limit of high sampling frequencies,
\begin{equation}
\label{eqn:kramers-moy}
D^{(k)}(\boldsymbol{x})=\lim\limits_{\tau\to\infty}\frac{1}{\tau}\left\langle \bigl[\boldsymbol{x}(t+\tau)-\boldsymbol{x}(t)\bigr]^k\big|x(t)=x\right\rangle\quad ,
\end{equation}
for $k=1$ and $k=2$,
respectively. $\boldsymbol{D^{(1)}}(\boldsymbol{x})$ is the
drift vector, while $D^{(2)}(\boldsymbol{x})$ exhibits the diffusion
matrix of the underlying process at position $\boldsymbol{x}$.
The limiting procedure (\ref{eqn:kramers-moy}) can be problematic
in case of a finite
time resolution of measured data. Moreover, any presence of
measurement or discretization noise seriously interferes 
with the convergence of the
limiting procedure \footnote{The application of this procedure 
in presence of measurement
noise  recently has been  investigated, see \cite{Boettcher06}.}. 

Recently, we proposed an iterative method that circumvents
this limiting procedure
\cite{Kleinhans05}. It is based on the minimization of the
Kullback-Leibler distance \cite{Haken:Information,Kullback} 
between the two time joint probability
distribution functions (pdf) obtained from the data and the simulated
process for a certain set of parameters, respectively. The 
starting configuration of this iterative procedure as well as a suitable
parametrization of drift and diffusion functions can be obtained by the
direct estimates based on the smallest reliable time increment $\tau$,
provided by (\ref{eqn:kramers-moy}).

On the other hand, the analysis of discrete stochastic processes by means of
\emph{maximum likelihood}-methods has made great progress in recent
years: Since it has become evident, that the maximisation of the
likelihood function is a powerful tool for the analysis of Markovian time series \cite{Lo88}, several methods have been proposed to optimise
the calculation of the required conditional transition pdfs
\cite{Sahalia02, Nicolau02}. For a recent study on the preferences of current
methods we refer to \cite{Hurn03}.

The intention of the present note is to derive a maximum likelihood
estimator for parameters of the parametrized drift vector and diffusion
matrix, that purely is based on the conditional and joint transition
pdfs of the dataset under consideration. By this means, the Kullback-Leibler estimator reappears in
case of an ensemble of individual measurements
-- but now physically well motivated. We want to point out, that the
evaluation of a specific parametrisation by means of the minimization
of the
Kullback-Leibler estimator yields great advantages, since this
function is bounded from below by the value $0$. Hence, the goodness
of a single parametrisation can be assessed.  

Moreover, with respect to our previous treatment
\cite{Kleinhans05}, a simplified maximum likelihood
estimator is introduced for the analysis of nonlinear time series exhibiting
Markovian properties. This estimator  leads to a
reasonable reduction of the required
computational effort compared to a direct application of the former
method and is proposed for future application in nonlinear time series
analysis.
Accurate results can be obtained even in the case of few or
sparsely sampled measurement data.
However, the relevance of the results obtained in the case of data sets
involving few data points carefully has to be
reconsidered  in a self consistent manner.

\section{Maximum Likelihood Estimation on Ensembles: Reconciliation with the
  Kullback-Leibler estimator \cite{Kleinhans05}}
We consider time series $\boldsymbol{x}(t_0),\ldots,\boldsymbol{x}(t_n)$, $t_i<t_{i+1}$ of $n$
recordings of a multivariate 
stochastic variable. Furthermore, we assume that the
time lag between consecutive observations is $\tau$. Henceforth, the abbreviation $\boldsymbol{x_i}:=\boldsymbol{x}(t_{0}+i\tau)$
will be used.

In this section, the estimation of drift and diffusion functions from an
ensemble of $N$ independent time series is considered. Such data sets generally are obtained from measurements on an ensemble of $N$ independent systems, that are performed simultaneously. In this vein, the time
evolution of the stochastic properties can be analysed. For the
present case, we restrict ourselves without
loss of generality
 to the analysis of  the first two consecutive measurements
$\boldsymbol{x_{0}^k}$ and $\boldsymbol{x_{1}^k}$ with $k\in[1,N]$ .

By means of the direct estimation described in \cite{Siegert98}, drift and
diffusion functions can be estimated from data  from the
Kramers-Moyal expansion coefficients (\ref{eqn:kramers-moy}). 
On the basis of this estimate,
models for the drift and diffusion function, respectively, can be
constructed, depending on a set of parameters, $\boldsymbol{A}$. This
procedure is described in greater detail in \cite{Kleinhans05}.

The likelihood of the current realization for one specific set of
parameters, $\boldsymbol{A}$, can be expressed by means of the joint pdf,
\begin{equation}
\label{eqn:en:jpdf}P(\boldsymbol{x_1^1},\boldsymbol{x_0^1},\boldsymbol{x_1^2},\boldsymbol{x_0^2},\ldots,\boldsymbol{x_1^N},\boldsymbol{x_0^N}|\boldsymbol{A})\quad.
\end{equation}
Since the individual $N$ processes are assumed to be
statistically independent of one another, this joint pdf degenerates into a
product of two point joint pdfs, 
\begin{equation}
\label{eqn:en:kette}P(\boldsymbol{x_1^1},\boldsymbol{x_0^1}|\boldsymbol{A})P(\boldsymbol{x_1^2},\boldsymbol{x_0^2}|\boldsymbol{A})\times\ldots\times
P(\boldsymbol{x_1^N},\boldsymbol{x_0^N}|\boldsymbol{A})\quad.
\end{equation}
This expression can be simplified considerably. 

First, we consider the
logarithm of (\ref{eqn:en:kette}), usually called 
log-likelihood function \cite{Kalbfleisch:II}, 
\begin{equation}
\label{eqn:en:logpdf}
\sum\limits_{k=1}^N \log\left[P(\boldsymbol{x_1^k},\boldsymbol{x_0^k}|\boldsymbol{A})\right]\quad.
\end{equation}
With help of $
  \hat{p}(\boldsymbol{x},\boldsymbol{x'}):=\frac{1}{N}\sum_{k=1}^{N}\delta(\boldsymbol{x}-\boldsymbol{x_1^k})\delta(\boldsymbol{x'}-\boldsymbol{x_{0}^k})
$
expression (\ref{eqn:en:logpdf}) finally can be evaluated by means of an integral, 
\begin{eqnarray}
\label{eqn:en:mle-integral}
&&\log\left[P(\boldsymbol{x_1^1},\boldsymbol{x_0^1},\boldsymbol{x_1^2},\boldsymbol{x_0^2},\ldots,\boldsymbol{x_1^N},\boldsymbol{x_0^N}|\boldsymbol{A})\right]\\
&&=N\int d\boldsymbol{x}\int d\boldsymbol{x'}\ \hat{p}(\boldsymbol{x},\boldsymbol{x'}|\tau) \log\left[P(\boldsymbol{x},\boldsymbol{x'}|\boldsymbol{A})\right]\quad.
\nonumber
\end{eqnarray}
Since the logarithm is a monotonically increasing function, 
the maximization of the
likelihood function is equivalent to the maximization of its
logarithm. The set $\boldsymbol{A}$, that maximizes the latter
expression, therefore forms the most likely
set of parameters under the current parametrization.
 
In \cite{Kleinhans05}, for the present case 
the \emph{min}imization of
the Kullback distance $\hat{K}[\boldsymbol{A}]$ of the
joint distributions has been proposed, 

\begin{subequations}
\label{eqn:en:kullback}
\begin{eqnarray}
  \hat{K}[\boldsymbol{A}]&=&\int d\boldsymbol{x}\int d\boldsymbol{x'}\ \hat{p}(\boldsymbol{x},\boldsymbol{x'})
  \log\left[\frac{\hat{p}(\boldsymbol{x},\boldsymbol{x'})}{P(\boldsymbol{x},\boldsymbol{x'}|\boldsymbol{A})}\right]\label{eqn:kullback:a}\\
  &=&\int d\boldsymbol{x}\int d\boldsymbol{x'}\ \hat{p}(\boldsymbol{x},\boldsymbol{x'})
  \log\left[\hat{p}(\boldsymbol{x},\boldsymbol{x'})\right]\label{eqn:en:kullback:b}\\
  &&-\int d\boldsymbol{x}\int d\boldsymbol{x'}\ \hat{p}(\boldsymbol{x},\boldsymbol{x'})
  \log\left[P(\boldsymbol{x},\boldsymbol{x'}|\boldsymbol{A})\right]\label{eqn:en:kullback:c}\quad .
\end{eqnarray}
\end{subequations}
The term (\ref{eqn:en:kullback:b}) is independent of the individual set
$\boldsymbol{A}$, while (\ref{eqn:en:kullback:c}) is conform to (\ref{eqn:en:mle-integral}). Therefore, minimization of
$\hat{K}[\boldsymbol{A}]$ evidently is equivalent to the maximization of the
likelihood of the set of parameters $\boldsymbol{A}$.

\section{\label{sect:timeseries}Maximum Likelihood estimation on Markovian time series}
Henceforth, individual time series
$\boldsymbol{x_0},\ldots,\boldsymbol{x_n}$ are considered. We assume that the
time lag between consecutive observations is $\tau$ and that the process
is stationary in a sense, that the statistical properties
are conserved during the
measurement period.

Let us further assume, that the data set under consideration exhibits Markovian
properties. This can be verified by means of the
Chapman-Kolmogorov equation \cite{Risken,Gardiner},
\begin{equation}
P(\boldsymbol{x_{i}}|\boldsymbol{x_{i-2}})=\int d\boldsymbol{x_{i-1}} P(\boldsymbol{x_{i}}|\boldsymbol{x_{i-1}})P(\boldsymbol{x_{i-1}}|\boldsymbol{x_{i-2}})\quad,
\end{equation}
that can be evaluated numerically. Although this condition is not
sufficient, it seems to be a very robust criterion. If Markovian properties
are not fulfilled, an increase of the number of observables by means of
a delay embedding of the data may help to
fulfil this constraint, if the amount of data is sufficiently high
for such an procedure \cite{Risken}.

If the process under consideration is ergodic, time averages can be
evaluated by means of ensemble averages. Then, also in this
case a reasonable parametrization and initial condition 
for the vector $\boldsymbol{A}$
can be obtained by the direct evaluation of
(\ref{eqn:kramers-moy}), as described in the previous section. Let us
now iterate the arguments of the previous section.

The likelihood of the current realization for a specific set of
parameters, $\boldsymbol{A}$, is
\begin{equation}
\label{eqn:jpdf}P(\boldsymbol{x_n},\ldots,\boldsymbol{x_0}|\boldsymbol{A})\quad.
\end{equation}
Since we assume Markov properties, this joint pdf degenerates into a
product of two point conditional pdfs, 
\begin{equation}
\label{eqn:markov-kette}P(\boldsymbol{x_n}|\boldsymbol{x_{n-1}},\boldsymbol{A})\times\ldots\times P(\boldsymbol{x_{1}}|\boldsymbol{x_{0}},\boldsymbol{A})
P(\boldsymbol{x_0}|\boldsymbol{A})\quad.
\end{equation}
This expression can be simplified by considering the
logarithm of the likelihood function. 
With the help of the definition 
\begin{equation}
  p(\boldsymbol{x},\boldsymbol{x'}):=\frac{1}{n}\sum_{i=1}^{n}\delta(\boldsymbol{x}-\boldsymbol{x_i})\delta(\boldsymbol{x'}-\boldsymbol{x_{i-1}})
\end{equation}
we finally obtain:
\begin{eqnarray}
\label{eqn:mle-integral}
\nonumber&&\log\left[P(\boldsymbol{x_n},\ldots,\boldsymbol{x_0}|\boldsymbol{A})\right]\\
&&=\log\left[
P(\boldsymbol{x_0}|\boldsymbol{A})\right]\\&&\hspace{.3cm} +n\int d\boldsymbol{x}\int d\boldsymbol{x'}\ p(\boldsymbol{x},\boldsymbol{x'}|\tau) \log\left[P(\boldsymbol{x}|\boldsymbol{x'},\boldsymbol{A})\right]\quad.
\nonumber
\end{eqnarray}
Following the maximum likelihood approach, this expression has to be
maximized with respect to $\boldsymbol{A}$. This is consistent with
the minimization of
\begin{eqnarray}\label{eqn:kullbackstrich}
  K'[\boldsymbol{A}]&=& -\frac{1}{n}\log\left[
P(\boldsymbol{x_0}|\boldsymbol{A})\right]\\&&
\int d\boldsymbol{x}\int d\boldsymbol{x'}\ p(\boldsymbol{x},\boldsymbol{x'})
  \log\left[\frac{p(\boldsymbol{x}|\boldsymbol{x'})}{P(\boldsymbol{x}|\boldsymbol{x'},\boldsymbol{A})}\right]\nonumber\quad.
\end{eqnarray}

It is obvious, that in the latter expression the first summand is
negligible for $n\gg 1$. Even in the case of smaller $n$, the first
measurement in some cases may not obey the stationary distribution
due to transient processes of the measurement. On the other hand, the
evaluation of the expression may be time-consuming since the stationary
distribution of the process is required. In conclusion, we propose to
solely perform the minimization of 
\begin{equation}
\label{eqn:kullback}
  K[\boldsymbol{A}]=\int d\boldsymbol{x}\int d\boldsymbol{x'}\ p(\boldsymbol{x},\boldsymbol{x'})
  \log\left[\frac{p(\boldsymbol{x}|\boldsymbol{x'})}{P(\boldsymbol{x}|\boldsymbol{x'},\boldsymbol{A})}\right]\quad .
\end{equation}

\section{Minimization Procedure for Drift-/Diffusion-Processes}
We would like to emphasize, that expression (\ref{eqn:kullback}) can be
evaluated numerically. It is a feature of drift and diffusion
processes, that the time evolution of the conditional pdf can be
obtained from the Fokker-Planck equation \cite{Risken}, 
\begin{eqnarray}
&\frac{\partial}{\partial
  t}P\left[\boldsymbol{x}(t)|\boldsymbol{x'}(t_{0})\right]=\biggl\{-\sum\limits_{i}\frac{\partial}{\partial x_i}D^{(1)}_{i}(\boldsymbol{x})&\\
&\hspace*{2cm}+\sum\limits_{i,j}\frac{\partial^2}{\partial x_i\partial x_j}D^{(2)}_{ij}(\boldsymbol{x}) \biggr\}P\left[\boldsymbol{x}(t)|\boldsymbol{x'}(t_{0})\right]\nonumber\quad.&
\end{eqnarray}
This equation can be treated efficiently by
implicit algorithms at least for the case
$\boldsymbol{x}\in\mathbb{R}$ and  $\boldsymbol{x}\in\mathbb{R}^2$, respectively
\cite{NrFortran}. Moreover, \emph{kernel density estimates} based
on numerical integration of the associated stochastic differential
equation can be applied, that are described in greater detail in \cite{Hurn03}.

The data under consideration can be reduced significantly by a
suitable discretization of data space in several bins. Typically, this
grid should coincidence with the spatial discretization required for
numerical solution of the Fokker-Planck equation. After discretization
and numerical evaluation of the expression
$P(\boldsymbol{x}|\boldsymbol{x'},\boldsymbol{A})$, equation
(\ref{eqn:kullback}) can be evaluated my means of a finite sum.

Eventually, the set $\boldsymbol{A}$, that minimizes
(\ref{eqn:kullback}) has to be investigated. This can be done by gradient 
method
or more efficient approaches
\cite{NrFortran}. We want to emphasize that in the 	majority of
cases a suitable starting value is obtained from the initial estimates (\ref{eqn:kramers-moy}). This is essential for a successful and fast minimization by any
numerical algorithm.

\section{Example}
In this section, the performance of the minimization procedure is
discussed by means of an example, that can be treated analytically. Further examples of
the minimization procedure are investigated in \cite{Kleinhans05,
  Nawroth07} for numerical and experimental data, respectively.

\begin{figure}
\includegraphics[width=\columnwidth]{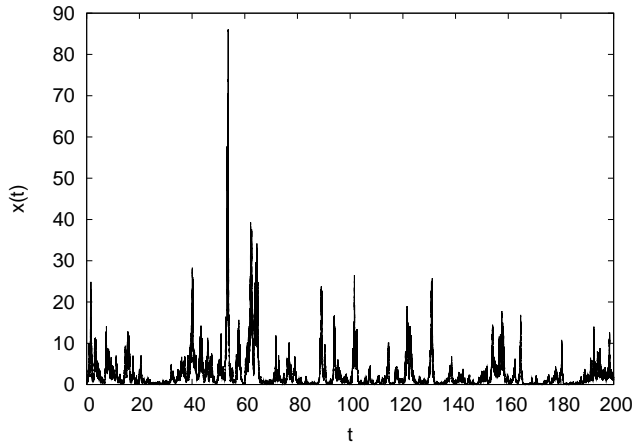}
\caption{\label{fig:sample}Detail of a sample process specified by equations
  (\ref{eqn:sample}) with $D=1.25$ and $\gamma=.75$.}
\end{figure}

\begin{subequations}
\label{eqn:sample}
We would like to address a drift and diffusion process in one
dimension with the diffusion term
\begin{equation}D^{(2)}(x)=D x^2\quad ,\end{equation} that in the long term limit obeys a
lognormal distribution
\begin{equation}p(x)=\frac{1}{x}\sqrt{\frac{1}{2\pi
      \frac{D}{\gamma}}}\exp\left\{-\frac{\left[\log\left(\frac{x}{x_0}\right)\right]^2}{2\frac{D}{\gamma}}\right\}\quad .\end{equation} 
This complies with the drift function
\begin{equation}
D^{(1)}(x)=x\left(D-\gamma\log\left(\frac{x}{x_0}\right)\right)\quad.
\end{equation} Thus, a stochastic process with a nonlinear drift
term is discussed, that is driven by multiplicative dynamical
noise. A feasible path of this process can be obtained by numerical
integration of the associated stochastic differential
equation. A sample graph of the process is exhibited in figure
\ref{fig:sample}. Thereby, It\^o's interpretation of stochastic
differential equations (sde) was
applied. 
For  the detailed properties of drift and diffusion processes we refer to
\cite{Risken}.
\end{subequations} 

For any one-dimensional process, the diffusion term significantly can
be simplified by means of a nonlinear transformation of the state variable:
For \begin{equation}y=y(x)=\int\limits_{x_0}^{x}dx'\
  \sqrt{\frac{D}{D^{(2)}(x')}}\quad,\end{equation}
the drift and diffusion functions transform to \cite{Risken}
\begin{subequations}\begin{eqnarray}
\tilde{D}^{(1)}(y)&=&\sqrt{\frac{D}{D^{(2)}(x(y))}}\\\nonumber&&\times\left[D^{(1)}(x(y))-\frac{1}{2}\frac{d
  D^{(2)}(x(y))}{dx}\right]\\
\tilde{D}^{(2)}(y)&=&D\quad.
\end{eqnarray}\end{subequations}
In the present case, the accordant nonlinear transformation and the
transformed drift and diffusion functions are:
\begin{subequations}
\begin{eqnarray}
y&=&\log(x)\\
\tilde{D}^{(1)}(y)&=&-\gamma y\\
\tilde{D}^{(2)}(y)&=&D
\quad .
\end{eqnarray}
\end{subequations}
Hence, the process is equivalent to an Ornstein-Uhlenbeck process in
the transformed variable $y$. For this process, the conditional
transition pdfs can be derived for finite time increment $\tau$ \cite{Risken}:
\begin{eqnarray}
\label{eqn:ou-cpdf}
\tilde{p}(y|y_0,\gamma,D)&=&\sqrt{\frac{1}{2\pi\frac{D}{\gamma}\left(1-e^{-2\gamma\tau}\right)}}\\\nonumber&&\times\exp\left[-\frac{\left(y-e^{-\gamma\tau}y_0\right)^2}{2\frac{D}{\gamma}\left(1-e^{-2\gamma\tau}\right)}\right]
\end{eqnarray}

Let us now consider the determination of the intrinsic parameters
$\gamma$ and $D$ from time series data. Imagine, the conditional pdfs
$p(x|x_0,\gamma_0,D_0)$ are known from time series data for a specific
set of parameters $(\gamma_0,D_0)$, that was applied for numerical
integration of the sde. This initial set will now be
reconstructed by means of the most likely set $(\gamma,D)$.
Following the argument of 
section \ref{sect:timeseries}, the minimization of
\begin{eqnarray}
K(\gamma,D)&=&\int\limits_{0}^\infty dx \int\limits_{0}^\infty dx_0\  p(x,x_0|\gamma_0,D_0)\\\nonumber&&\times\log\left[\frac{p(x|x_0,\gamma_0,D_0)}{P(x|x_0,\gamma,D)}\right]
\end{eqnarray}
is sufficient for this purpose.

This expression can be calculated by means of the underlying
Ornstein-Uhlenbeck process,
\begin{eqnarray}
K(\gamma,D)&=&\int\limits_{-\infty}^\infty dy
\int\limits_{-\infty}^\infty dy_0\
\tilde{p}(y,y_0|\gamma_0,D_0)\\\nonumber&&\times\log\left[\frac{\tilde{p}(y|y_0,\gamma_0,D_0)}{\tilde{P}(y|y_0,\gamma,D)}\right]\quad.\end{eqnarray}

As a first step, the logarithm can be evaluated for the specific
conditional transition pdfs of the Orstein-Uhlenbeck process under
consideration, (\ref{eqn:ou-cpdf}). It turns out, that the
$K(\gamma,D)$ solely is determined
by the second order moments of $y$ and $y_0$:

\begin{widetext}\begin{eqnarray}
\nonumber K(\gamma,D)&=&\left\langle
  y^2\right\rangle_{y,y_0}\left(\frac{1}{2\frac{D}{\gamma}\left(1-e^{-2\gamma\tau}\right)}-\frac{1}{2\frac{D_0}{\gamma_0}\left(1-e^{-2\gamma_0\tau}\right)}\right)+
\left\langle  y_0^2\right\rangle_{y,y_0}\left(\frac{e^{-2\gamma\tau}}{2\frac{D}{\gamma}\left(1-e^{-2\gamma\tau}\right)}-\frac{e^{-2\gamma_0\tau}}{2\frac{D_0}{\gamma_0}\left(1-e^{-2\gamma_0\tau}\right)}\right)\\&&+ \left\langle y
   y_0\right\rangle_{y,y_0}\left(-\frac{e^{-\gamma\tau}}{\frac{D}{\gamma}\left(1-e^{-2\gamma\tau}\right)}+\frac{e^{-\gamma_0\tau}}{\frac{D_0}{\gamma_0}\left(1-e^{-2\gamma_0\tau}\right)}\right)+\frac{1}{2}\log\left[\frac{\frac{D}{\gamma}\left(1-e^{2\gamma\tau}\right)}{\frac{D_0}{\gamma_0}\left(1-e^{2\gamma_0\tau}\right)}\right]
\end{eqnarray}\end{widetext}

\begin{figure}
\includegraphics[width=\columnwidth]{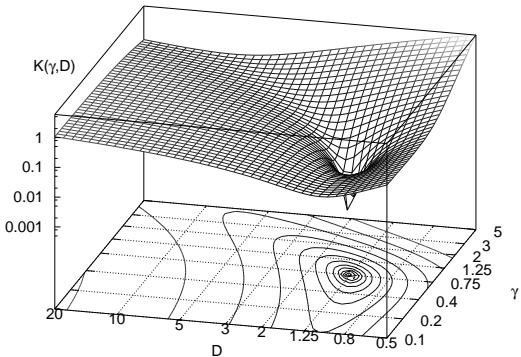}
\caption{\label{fig:kullback}Kullback distance between processes
  $(\gamma_0=0.75,D_0=1.25)$ and $(\gamma,D)$ as function of $\gamma$ and
  $D$ for $\tau=0.5$. A distinct minimum at $(\gamma,D)=(\gamma_0,D_0)$ is
  evident. The contour lines are located at $z=2^i$ for
  $i=-11,\ldots,0$. For the sake of clearness, the z-axis is scaled logarithmically.}
\end{figure}
Finally, a closed form for the function $K(\gamma,D)$
can be derived:
\begin{eqnarray}
K(\gamma,D)&=&\frac{1}{2}\log\left[\frac{\frac{D}{\gamma}\left(1-e^{2\gamma\tau}\right)}{\frac{D_0}{\gamma_0}\left(1-e^{2\gamma_0\tau}\right)}\right]\\
&&+\frac{\frac{D_0}{\gamma_0}}{\frac{D}{\gamma}}\frac{1+e^{2\gamma\tau}-2e^{(\gamma-\gamma_0)\tau}}{2\left(e^{2\gamma\tau}-1\right)}-\frac{1}{2}\nonumber
\end{eqnarray}

For the initial set $(\gamma_0,D_0)=(0.75,1.25)$, this function is exhibited
in figure \ref{fig:kullback}. A distinct minimum at
$(\gamma,D)=(0.75,1.25)$ is evident, that complies with the initial set of parameters. In case of an application of
the minimization procedure to real measurements, this minimum would
have to be approached by means of gradient methods or advanced
minimization algorithms \cite{NrFortran}.

\section{Conclusion}
In conclusion, the likelihood functions of stochastic processes have
been derived for two specific cases. First, ensembles of measurements
on these processes were considered. In this connection, the
iterative procedure proposed in \cite{Kleinhans05} has been approved
and physically motivated.

Moreover, the maximum likelihood approach has been adapted to the
needs of non-linear time series analysis. For the case of Markovian
processes, an integral form of the estimator has been derived. A slight
simplification of this estimator, equation (\ref{eqn:kullback}), is
purely based on two point conditional pdfs, that can be calculated numerically from the Fokker-Planck equation in case of
drift and diffusion processes. The integral form of the estimator
allows for the reduction of huge datasets to their conditional
transition pdfs prior to the iterative analysis procedure.

Finally, the meaning of the
optimal set of parameters, $\boldsymbol{A}$, that is obtained by
application of the method described in \cite{Kleinhans05},
has been made explicit on the basis of the maximum likelihood approach:
 It is the most likely set of parameters with
respect to the current parametrization. As a consequence,
the proposed procedure can be applied even to time series 
that suffer from sparse
data points and that could not safely be processed by the former
methods without this knowledge.


\end{document}